\documentclass[12pt,a4paper]{article}
\usepackage{cmap}
\usepackage[T2A]{fontenc}
\usepackage[cp1251]{inputenc}
\usepackage[english, russian]{babel}
\usepackage{amsfonts,amsmath,amssymb,amsthm,longtable,hhline}
\usepackage{bm}
\usepackage{cite}
\usepackage{color,soul}
\usepackage{color,xcolor}
\usepackage{multicol}
\usepackage{tabularx,caption,multirow}
\usepackage{ulem}
\usepackage{titlesec}

\makeatletter
\renewcommand{\@biblabel}[1]{#1.} 
\makeatother
\makeatletter

\newcommand{\clh}[1]{\colorbox{yellow}{#1}}%
\newcommand{\clhp}[1]{\colorbox{yellow}{\parbox{\textwidth}{#1}}}%
\newlength\aptextwidth
\definecolor{BrickRed}{rgb}{0.588,0.098,0.055}

\def\app#1{\textcolor{red}{\textrm{#1}}}

\def\noblue#1{\ifmmode \text{#1}\else #1\fi}
\def\noclh#1{\ifmmode \text{#1}\else #1\fi}
\def\rem#1{}

\let\le=\leqslant

\let\ge=\geqslant

\def\ttable#1. #2{\begin{table}[t]\tablehat{#1}{#2}}
\def\mtable#1. #2{\begin{table}[hbtp]\tablehat{#1}{#2}}
\def\ptable#1. #2{\begin{table}[p]\tablehat{#1}{#2}}
\def\tablehat#1#2{\centering\small \vbox{\parindent=0pt
  \leftskip=0pt plus.5\hsize \rightskip=\leftskip \parfillskip=0pt
  ТАБЛИЦА #1\\ #2}\nobreak\medskip\medskip }
\def\texendtable{\end{table}}
\def\notation{\par\ifnum\lastpenalty<25000 \bigbreak \fi
  \noindent\triangle\enspace\ignorespaces}

\def\hline{\noalign{\hrule}}

\let\ts=\textstyle

\let\bl=\bigl \let\br=\bigr
 
\let\BL=\biggl \let\BR=\biggr

\def\eqnitemskip{\ifhmode \else \par
  \ifnum\lastpenalty>24999
    \ifnum\lastpenalty=25004 \fi
  \else \medbreak \fi \fi }
\def\eqnitem #1. {\eqnitemskip
  {\setbox0=\hbox{$#1^\circ$.\enspace}%
  \ifdim\wd0>\parindent \box0\ignorespaces \else
  \hbox to\parindent{\unhbox0\hss}\ignorespaces\fi}}
\def\eqnitemnobreak #1. {\noindent
  {\setbox0=\hbox{$#1^\circ$.\enspace}%
  \ifdim\wd0>\parindent \box0\ignorespaces \else
  \hbox to\parindent{\unhbox0\hss}\ignorespaces\fi}}
\newdimen\eqnparindent
\eqnparindent=\parindent
\def\eqnitem #1. {\eqnitemskip\noindent\hskip\eqnparindent $#1^\circ$.\enspace\ignorespaces }
\def\eqnitemnobreak #1. {\noindent\hskip\eqnparindent $#1^\circ$.\enspace\ignorespaces }
\def\simpleitem #1. {\eqnitemskip\noindent\hskip\eqnparindent #1.\enspace\ignorespaces }

\def\eqalignno#1{\displ@y \tabskip\centering
  \halign to\displaywidth{\hfil$\@lign\displaystyle{##}$\tabskip\z@skip
    &$\@lign\displaystyle{{}##}$\hfil\tabskip\centering
    &\llap{$\@lign\eqnofont##$}\tabskip\z@skip\crcr
    #1\crcr}}
\let\eqalignno=\eqalignm
\def\eqcenter#1{\displ@y \tabskip\centering
  \halign{\hfil$\displaystyle{##}$\hfil\crcr
    #1\crcr}}
\def\eqcenterno#1{\displ@y \tabskip\centering
  \halign to\displaywidth{\hfil$\@lign\displaystyle{##}$\hfil
    \tabskip\centering&\llap{$\@lign\eqnofont##$}\tabskip\z@skip\crcr
    #1\crcr}}
\def\texcases#1{\left\{\,\vcenter{\normalbaselines\m@th
    \ialign{$##\hfil$&\quad##\hfil\crcr#1\crcr}}\right.}
\def\Displaylines#1{\vcenter{\displ@y \tabskip\z@skip
  \halign{\hbox to\displaywidth{$\@lign\hfil\displaystyle##\hfil$}\crcr
    #1\crcr}}}

\def\tan{\mathop{\operator@font tg}\nolimits}

\def\arb{is an arbitrary constant}
\def\arbs{are arbitrary constants}

\begin{document}
\large 

\bigskip
\centerline{\bf\Large Two-dimensional nonlinear Schr\"odinger equations}
\centerline{\bf\Large with potential and dispersion given by arbitrary functions:}
\centerline{\bf\Large Reductions and exact solutions\clh{$^{*}$}}

\bigskip
\medskip

\centerline{Andrei D. Polyanin}
 \medskip
\centerline{Ishlinsky Institute for Problems in Mechanics, Russian Academy of Sciences,}
\centerline{pr. Vernadskogo 101, bldg. 1, Moscow, 119526 Russia}
\smallskip
\centerline{e-mail: polyanin@ipmnet.ru}
\bigskip
\let\thefootnote\relax\footnotetext{
\hskip-20pt\clhp{$^*$ This article was published in the \textbf{Open Journal of Mathematical Sciences, 2026, Vol.~10, pp.~173--194};
doi:\! 10.30538/oms2026.0279.}}

\vspace{5.0ex}

{For the first time, a nonlinear Schr\"odinger equation of the general form is considered, depending on time and two spatial variables, the potential and dispersion of which are specified by two arbitrary functions.
This equation naturally generalizes a number of simpler} nonlinear partial differential equations {encountered in various fields} of theoretical physics, including nonlinear optics, superconductivity, and plasma physics.
Two- and one-dimensional reductions are described, which reduce the studied nonlinear Schr\"odinger equation to simpler equations of lower dimension or ordinary differential equations (or systems of ordinary differential equations).
{In addition to the general Schr\"odinger equation with two arbitrary functions, related nonlinear partial differential equations are also examined, in which the dispersion function is specified arbitrarily while the potential function is expressed in terms of it.
For all considered classes of nonlinear PDEs, using the methods of generalized and functional separation of variables, as well as the semi-inverse approach and the principle of structural analogy of solutions, many new exact solutions have been found, which are expressed in terms of elementary or special functions,
or in the form of quadratures.
Both Cartesian and polar coordinate systems are employed to analyze the equations under consideration.}
Special attention is paid to finding solutions with radial symmetry.
{It is shown that the nonlinear Schr\"odinger equation, in which the functions defining the potential and dispersion are linearly related (one of these functions can be chosen arbitrarily), can be reduced to a two-dimensional nonlinear PDE that admits exact linearization.}
The exact solutions obtained in this work can be used as test problems intended {for verifying the adequacy and assessing} the accuracy of numerical and approximate analytical methods for solving complex nonlinear PDEs of mathematical physics.
\medskip


\section{Introduction}\label{sec1}

\subsection{Classical nonlinear Schr\"odinger equation and related nonlinear PDEs}

In a number of branches of theoretical physics, one encounters the classical one-dimensional nonlinear Schr\"odin\-ger equation with cubic nonlinearity, which is written as follows \cite{Ag1, Ag2, Ag3, Ag4, Ag5, Ag6}:
\begin{equation}
iw_t+w_{xx}+k|w|^2w=0,
\label{Schrodinger-eq}
\end{equation}
where $w=w(x,t)$ is the desired complex-valued function of real arguments,
$t$ is the time variable, $x$ is the spatial variable, $k$  is a parameter of the equation, $i^2=-1$.

The theoretical and experimental justification of Eq. \eqref{Schrodinger-eq} in nonlinear optics is the subject of the works \cite{UFN,Has1,Has2,Has3}.
Eq.~\eqref{Schrodinger-eq} belongs to the class of integrable PDEs;
it has an infinite number of conservation laws and passes the Painlev\'e test \cite{Ag4, Ag5, Pain3, Pain4, Pain5}.
The Cauchy problem for this equation can be solved by the inverse scattering method \cite{Ag4, Ag5}.
Exact solutions of the classical nonlinear Schr\"odinger equation \eqref{Schrodinger-eq} are given, for example, in \cite{polzai2012,kha2019,pol2025}.

Related nonlinear partial differential equations of the form
\begin{align}
iw_t+w_{xx}+g(|w|)w&=0,\label{Schrodinger-eq2}\\
iw_t+[f(|w|)w]_{xx}+g(|w|)w&=0,\label{Schrodinger-eq3}
\end{align}
and many other one-dimensional nonlinear Schr\"odinger-type equations that appear in the literature can be found, for example, in
\cite{polzai2012,kha2019,pol2025,Kudr3,Kudr5, Kudr6,Bis1,Bis3,Bis4,Bis5,Bis6,izg2023,akr2023,polkud2,polkud2a,polkud3,polkud4,yao2025}
(it should be noted that the last three works \cite{polkud3,polkud4,yao2025} considered nonlinear Schrödingers equations complicated by delay).
In nonlinear optics, the potential $g(|w|)$ in equations \eqref{Schrodinger-eq2} and \eqref{Schrodinger-eq3} characterizes the law of interaction of a light pulse with the fiber material. The function $f(|w|)$ in equation \eqref{Schrodinger-eq3} describes nonlinear dispersion.

Exact solutions of Eq. \eqref{Schrodinger-eq2} for the power potential $g(|w|)=a|w|^k$ were considered, for example, in \cite{polzai2012,kha2019,pol2025}.
Eq. \eqref{Schrodinger-eq2} with $g(|w|)=a[1-\exp(k|w|)]$ occurs in laser physics and plasma theory (see, for example, \cite{bol1978}).
Solutions with generalized separation of variables of the nonlinear PDE \eqref{Schrodinger-eq2} with an arbitrary function $g(|w|)$ are given in \cite{polzai2012,pol2025}. Exact solutions and reductions of the nonlinear Schr\"odinger equation of general form \eqref{Schrodinger-eq3} with two arbitrary functions $f(|w|)$ and $g(|w|)$ are described in \cite{polkud2}.

Analytical and numerical mathematical models described by nonlinear two-dimensional Schr\"odinger equations with parameters have been studied in a number of papers. The paper \cite{Tw_1} explores the stochastic (2+1)-dimensional cubic-quartic nonlinear Schr\"odinger equation with non-local law and spatio-temporal dispersion.  The (2+1)-dimensional nonlinear Schr\"odinger equation with fourth-order nonlinearity and dispersion has been studied in \cite{Tw_2}.
Some different solutions in the form of bright and dark optical modulated solitons have been derived using different approaches. The (2+1)-dimensional chiral nonlinear Schr\"odinger equation via two random sources has been investigated in paper \cite{Tw_3}, where the equation forced by multiplicative noise in Ito sense and the spatio-temporal coefficient has been considered. In work \cite{Tw_4} the nonlinear Schr\"odinger equation in $2+1$ dimensions with arbitrary nonlinearity exponent in the presence of an external confining potential has been presented.
The (2+1)-dimensional chiral nonlinear Schr\"odinger equation has been presented the edge states of the Hall effect in paper \cite{Tw_5}, where the authors have been used the bifurcation conditions determined using the theory of planar dynamical systems and the phase pictures of the system have been produced using quantitative analysis.
The (3+1)-dimensional nonlinear hyperbolic Schr\"odinger model has been applied to study water wave development, soliton dynamics and electromagnetic field propagation in paper \cite{Tw_6}, where the chaotic structure for the governing model has been examined with and without disturbance using chaos theory and bifurcation theory.
The nonlinear Schr\"odinger equation in (2+1)-dimensions has been addressed influenced by cubic-quintic-septic law nonlinearity and spatial dissipations effects in work \cite{Tw_7}.
The (2+1)-dimensional nonlinear Schr\"odinger equation with fourth-order nonlinearity and dispersion has been used investigated in paper \cite{Tw_8}, where several optical solitons and other traveling wave solutions have been discovered using the modified extended direct algebraic method.
The (2+1)-dimensional stochastic chiral nonlinear Schr\"odinger equation has been studied in paper \cite{Tw_11}, where a number of novel solutions such as periodic, singular, dark and bright solitons solutions have been retrieved.
The (2+1)-dimensional nonlinear Schr\"odinger equation has been investigated in paper \cite{Tw_12} which characterizes the transmission of optical pulses through optical fibers exhibiting refractive index variations corresponding to light intensity changes.
The Bernoulli sub-equation function method has been used to construct new exact travelling wave solutions for two important physical models in paper \cite{Tw_13}: (2+1)-dimensional hyperbolic nonlinear Schr\"odinger equation and (2+1)-dimensional Heisenberg ferromagnetic spin chain equation.
The Lie symmetry analysis method has been applied to time fractional coupled (2+1)-dimensional nonlinear Schr\"odinger equations in paper \cite{Tw_14}.
The approach has been utilized to construct several novel optical soliton solutions for the generalized integrable (2+1)-dimensional nonlinear Schr\"odinger system with conformable derivative in paper \cite{Tw_17}, where the dynamics of bifurcation behavior and chaos analysis in the system has been investigated.
Some solitary solutions of (2+1)-dimensional variable coefficient nonlinear Schr\"odinger equations have been obtained through computerized symbolic computation in \cite{Tw_19} by applying the $F$-expansion and homogeneous balance method.
The existence and stability of nonlinearly chirped periodic waves and soliton structures in an optical medium wherein the pulse propagation have been  governed by the (1+2)-dimensional chiral nonlinear Schr\"odinger equation in paper \cite{Tw_22}.
Bilinear B\"acklund transformation and optical soliton solutions to a (3+1)-dimensional coupled nonlinear Schr\"odinger equation with variable coefficients
were considered in \cite{hao2025}.

\subsection{Terminology: what is meant by exact solutions of nonlinear PDEs}

As in \cite{polzhu2022}, in this paper, exact solutions of partial differential equations are
understood to mean the following solutions:
\goodbreak

(\textit{a})\enspace Solutions that are expressed in terms of elementary functions.

(\textit{b})\enspace Solutions that are expressed in quadratures, i.e. through elementary functions, functions included in the equation (this is necessary if the equation contains arbitrary or special functions) and indefinite integrals.

(\textit{c})\enspace Solutions that are expressed in terms of solutions of ordinary differential equations (ODEs) or finite systems of such equations.

Various combinations of the solutions described in paragraphs (\textit{a})--(\textit{c}) are also allowed. In cases (\textit{a}) and (\textit{b}) the exact solution can be presented in explicit, implicit, or parametric form.

It is important to note that exact solutions are a kind of mathematical standards that are often used as test problems to check the adequacy and assess the accuracy of numerical and approximate analytical methods for integrating nonlinear partial differential equations.

Reductions of a differential equation under consideration are usually understood to be equations of lower dimension or lower order, all solutions of which are solutions of the given equation. Reductions lead to simpler equations and play a key role in constructing exact solutions of nonlinear PDEs.

In this paper, various modifications of the methods of generalized and functional separation of variables \cite{polzai2012,polzhu2022,galsvi2007}, {semi-inverse approach, and the principle of structural analogy of solutions \cite{polkud3,aks2021,pol2026}} were used to find exact solutions and reductions of two-dimensional spatial generalizations of nonlinear equations \eqref{Schrodinger-eq2} and \eqref{Schrodinger-eq3}, which contain one or two arbitrary functions.

\section{Nonlinear Schr\"odinger equations with two spatial variables}

\subsection{Nonlinear Schr\"odinger equations under consideration and their properties. Preliminary description of the semi-inverse approach}

$1^\circ$. In this paper we will consider two-dimensional nonlinear Schr\"odinger equations of the general form
\begin{align}
iw_t+\Delta w+g(|w|)w&=0,\label{eq00a}\\
iw_t+\Delta[f(|w|)w]+g(|w|)w&=0,\label{eq00b}
\end{align}
where $w=w(x,y,t)$ is the desired complex-valued function of real arguments, $t$ is the time variable,
$x$ and $y$ are the Cartesian coordinates, $\Delta$ is the Laplace operator (i.e.
$\Delta w=w_{xx}+w_{yy}$), and $i^2=-1$.
Equations \eqref{eq00a} and \eqref{eq00b} are two-dimensional generalizations of equations \eqref{Schrodinger-eq2} and \eqref{Schrodinger-eq3}. The real functions {$f(r)$ and $g(r)$ with $r=|w|$} included in these nonlinear PDEs and characterizing the dispersion and potential will be considered {rather arbitrary
(it is assumed that the function $f$ is twice continuously differentiable, and the function $g$ is continuous).}

A group classification of equation \eqref{eq00a} {with an arbitrary potential $g$ was carried out in \cite{bar1991}.}
Some exact solutions of this equation {with two, three and $n$ spatial variables} were obtained in {\cite{polzai2012,bar1991}}.

{It is important to note that the nonlinear Schr\"odinger equation with two spatial variables \eqref{eq00b}, which includes two arbitrary functions $f$ and $g$, has not been considered before. Therefore, all the exact solutions of this equation obtained in the following sections are new.}

{$2^\circ$. In addition to the general Schr\"odinger equation with two arbitrary functions \eqref{eq00b}, we will also consider equations of the form \eqref{eq00b} in which the function $f(r)$ will be set arbitrarily, and the second function $g(r)$ will be set parametrically by two relations
\begin{equation}
\begin{aligned}
rf(r)&=h(\xi),\\
g(r)&=M(\xi,r,h,h'_\xi,h''_{\xi\xi}),
\end{aligned}
\label{eqxx}
\end{equation}
where $M$ is some given function of its arguments (it is derived from an ODE obtained as a result of a reduction),
$\xi$ is a parameter (which can be a spatial variable or some ansatz of independent variables),
$h=h(\xi)$ is a twice continuously differentiable function, the form of which can vary depending on a priori ideas about the structure of the solution.
By inserting a specific invertible function $h(\xi)$ into \eqref{eqxx}, after eliminating $\xi$ we obtain a representation of the potential function in the form $g(r)=Q(r,f(r))$. In this case, the function $f(r)$ remains arbitrary, and the function $g(r)$ is expressed through it (further, using specific examples, it will be shown that the invertibility of the function $h(\xi)$ included in \eqref{eqxx} is not a necessary condition for such a representation).}

{By specifying in \eqref{eqxx} various concrete functions $h=h(\xi)$, we will obtain various classes of nonlinear Schr\"odinger equations depending on one arbitrary function $f(r)$. In particular, substituting $h(\xi)=\xi$ in \eqref{eqxx}, we get $g(r)=M(rf(r),r,rf(r),1,0)$.}

{The semi-inverse approach described above, in which one function of the equation under consideration is specified arbitrarily, and the second is expressed through it using relations \eqref{eqxx}, will be discussed in more detail
in Section \ref{ss:3.2}, and specific examples of its application will be given in Sections \ref{ss:3.3}, \ref{ss:3.4}, and \ref{ss:4.3}).
It should be noted that the exact solutions to nonlinear Schr\"odinger PDEs of the form \eqref{eq00b}, obtained further in this article, are new.}

\textit{$3^\circ$. Properties of equations} \eqref{eq00a} and \eqref{eq00b}.
Let $\app{$w$}(x,y,t)$ be a solution of equation \eqref{eq00b}. Then the functions
\begin{equation}
\begin{aligned}
w_1&=\exp(-iA)w(\pm x+C_1,\pm y+C_2,t+C_3),\\
w_2&=w(x\cos\beta-y\sin\beta,x\sin\beta+y\cos\beta,t),
\end{aligned}
\label{eq00c}
\end{equation}
where $A$, $C_1$, $C_2$, $C_3$, and $\beta$ are arbitrary real constants, are also solutions of this equation
(the plus or minus signs in $w_1$ are chosen arbitrarily).
For the simpler equation \eqref{eq00a}, the following property is added to the properties of \eqref{eq00c}:
if $w(x,y,t)$ is a solution to the equation \eqref{eq00a}, then the function
\begin{equation}
w_3=\exp\bl\{-i[\lambda_1x+\lambda_2y+(\lambda_1^2+\lambda_2^2)t]\br\}w(x+2\lambda_1t,y+2\lambda_2t,t),
\label{eq00cc}
\end{equation}
where $\lambda_1$ and $\lambda_2$ are arbitrary real constants, is also a solution of this equation.

The formulas \eqref{eq00c} and \eqref{eq00cc}, containing seven free parameters, allow using simpler partial solutions of the equations \eqref{eq00a} and \eqref{eq00b} to construct more complex exact solutions of these equations. In particular, the formula \eqref{eq00cc} makes it possible to obtain some non-stationary solutions of the equation \eqref{eq00a} using its stationary solutions.

{\textit{Remark 1}.
Symmetries and exact solutions of the nonlinear Schr\"odinger equation \eqref{eq00a} with three spatial variables with a potential of the special form $g(|w|)=a+b|w|^2+c|w|^4$ were considered in \cite{gag1988,gag1989,ibr1995}
(for exact solutions of related PDEs with some given specific functions $g$, see also \cite{polzai2012,bar1991,ibr1995,kha2019,zho2008,sea2012}).}

\subsection{Transformation of the nonlinear Schr\"odinger equation to a system of two real PDEs}

For subsequent analysis, it is convenient to represent the desired {complex-valued} function in exponential form
\begin{equation}
w=re^{i\varphi},\quad \ r=|w|,
\label{eq02}
\end{equation}
where $r=r(x,y,t)\ge 0$ and $\varphi=\varphi(x,y,t)$ are real functions.

{Using relation \eqref{eq02}, we have
\begin{equation}
f(|w|)w=f(r)re^{i\varphi}=he^{i\varphi},\quad \  h=rf(r),
\label{eq02a}
\end{equation}}
Differentiating \eqref{eq02} {and \eqref{eq02a}, we successively} obtain
\begin{equation}
\begin{aligned}
w_t&=(r_t+ir\varphi_t)e^{i\varphi},\\
[f(|w|)w]_x&=(he^{i\varphi})_x=(h_x+ih\varphi_x)e^{i\varphi},\quad [f(|w|)w]_y=(h_y+ih\varphi_y)e^{i\varphi},\\
[f(|w|)w]_{xx}&=[(h_x+ih\varphi_x)e^{i\varphi}]_x=[h_{xx}-h\varphi_x^2+i(2h_x\varphi_x+h\varphi_{xx})]e^{i\varphi},\\
[f(|w|)w]_{yy}&=[(h_y+ih\varphi_y)e^{i\varphi}]_x=[h_{yy}-h\varphi_y^2+i(2h_y\varphi_y+h\varphi_{yy})]e^{i\varphi},\\
\Delta[f(|w|)w]&=[\Delta h-h|\nabla\varphi|^2+i(2\nabla h\cdot\nabla\varphi+h\Delta\varphi)]e^{i\varphi},\quad h=rf(r),
\end{aligned}
\label{eq03}
\end{equation}
where the following notations are used
\begin{equation*}
\Delta h=h_{xx}+h_{yy},\quad \
|\nabla\varphi|^2=\varphi_x^2+\varphi_y^2,\quad \ \nabla h\cdot\nabla\varphi=h_x\varphi_x+h_y\varphi_y.
\end{equation*}

Now we substitute \eqref{eq03} into \eqref{eq00b}, and then divide {all terms} by $e^{i\varphi}$.
Further equating the real and imaginary parts of the obtained relation to zero, we arrive at the following system of two real partial differential equations:
\begin{equation}
\begin{aligned}
-r\varphi_t+h_{xx}+h_{yy}-h(\varphi_x^2+\varphi_y^2)+rg(r)&=0,\\
r_t+2(h_x\varphi_x+h_y\varphi_y)+h(\varphi_{xx}+\varphi_{yy})&=0,\quad \ h=rf(r).
\end{aligned}
\label{eq04}
\end{equation}

Thus, we have shown that the complex-valued two-dimensional nonlinear\break
Schr\"odinger equation \eqref{eq00b} by introducing two auxiliary functions $r$ and $\varphi$ according to the formula \eqref{eq02} is transformed to a system of two real PDEs \eqref{eq04}. A simpler nonlinear equation \eqref{eq00a} in this case is reduced to a system of PDEs \eqref{eq04}, in which we must set $f(r)\equiv 1$, which gives $h=r$.

The representation \eqref{eq02} and the system of PDEs \eqref{eq04} will be used further to construct exact solutions of the nonlinear equations \eqref{eq00a} and \eqref{eq00b}.

\textit{Remark {2}}.
The second equation in system \eqref{eq04} can be represented in divergent form as a conservation law:
$$
H_t+(h^2\varphi_x)_x+(h^2\varphi_y)_y=0,\quad \ H=\int h(r)\,dr.
$$

\section{Solutions and reductions of the nonlinear Schr\"odinger equation in the Cartesian coordinate system}

Below we describe some exact solutions and reductions of the general two-dimensional nonlinear Schr\"odinger equation \eqref{eq00b}, which contains two arbitrary functions $f(z)$ and $g(z)$. To construct these solutions, we use a representation of the solution in exponential form \eqref{eq02} and a system of two real PDEs written in a Cartesian coordinate system \eqref{eq04}.

\subsection{Two-dimensional traveling wave type solutions with constant amplitude}

System \eqref{eq04} has a simple exact solution of the two-dimensional traveling wave type
\begin{equation}
\begin{aligned}
r&=A,\quad \ \varphi=Bt+C_1x+C_2y+C_3,\\
B&=g(A)-f(A)(C_1^2+C_2^2),
\label{eq05ab}
\end{aligned}
\end{equation}
where $A$, $C_1$, $C_2$, and $C_3$ are arbitrary constants (from here on all arbitrary constants included in the solutions are considered real, which is not specifically stated). Formulas \eqref{eq02} and \eqref{eq05ab} define a solution of the nonlinear Schr\"odinger equation \eqref{eq00b} that is periodic in both time $t$ and both spatial variables $x$ and~$y$.

\subsection{Closed-form solution, periodic in time and in one spatial variable, with amplitude dependent on the other spatial variable}\label{ss:3.2}

System \eqref{eq04} admits a periodic in time $t$ and spatial coordinate $y$ exact solution with variable amplitude
\begin{equation}
r=r(x),\quad \ \varphi=At+By+C,
\label{new01}
\end{equation}
where $A$, {$B$}, and $C$ are arbitrary constants, and the function $r=r(x)$ is described by the nonlinear autonomous ODE
\begin{equation}
h''_{xx}-B^2h-Ar+rg(r)=0,\quad \ h=rf(r).
\label{new02}
\end{equation}

{In what follows, it is assumed throughout that the potential function $f=f(r)$ satisfies the following conditions:
\begin{equation}
f\ge 0,\quad \ f'_r\ge 0,\quad \ f\not\equiv 0\quad \ \text{for} \ \ r\ge 0.
\label{eq600}
\end{equation}}

{Let us now multiply all terms of equation \eqref{new02} by $2h'_x$, and then integrate with respect to $x$.
Taking into account the relations $2h'_xh''_{xx}=[(h'_x)^2]'_x$ and $2hh'_x=(h^2)'_x$, as a result we obtain}
\begin{equation}
(h'_x)^2=B^2h^2+2\int [Ar-rg(r)]dh+C_1,\quad \ h=rf(r),
\label{eq06*****}
\end{equation}
{where $C_1$ \arb.
Since the function $h$ must be real, the right-hand side of the relation \eqref{eq06*****} must be nonnegative, which leads to the condition
$$
B^2h^2+2\int [Ar-rg(r)]dh+C_1\ge 0.
$$}

{The first integral~\eqref{eq06*****},} after resolution with respect to the derivative $h'_x$, is reduced to a first-order ODE with separable variables.
Integrating this equation, we obtain the general solution of equation \eqref{new02} in implicit form
\begin{equation}
\int\BL\{B^2h^2+2\int [Ar-rg(r)]dh+C_1\BR\}^{\!-1/2}dh=C_2\pm x,
\label{eq06******}
\end{equation}
where $C_1$ and $C_2$ are arbitrary constants, $h=rf(r)$, and $dh=[f(r)+rf'(r)]\,dr$.

The general closed-form solution of the ODE \eqref{new02}, represented in quadratures in the implicit form \eqref{eq06******},
is difficult to study and utilize due to the presence of two arbitrary functions $f(r)$ and $g(r)$, which are included in the integrands.

{Note that for the defining functions of power-law form
$$
f(r)=\beta r^\sigma ,\quad \ g(r)=\gamma  r^\sigma ,
$$
where $\beta >0$, $\gamma  >0$, and $\sigma >0$ are free parameters, solution \eqref{eq06******} with $A>0$, $C_1=0$, and $\beta B^2 > \gamma  $ can be expressed explicitly in terms of elementary functions by the following formulas:
\begin{align*}
r(x)&=\left[\frac{2A(\sigma +1)}{(\sigma +2)(\beta B^2-\gamma)}\sinh^{2}\left( \frac{\sigma \sqrt{\beta B^2-\gamma}}
{2\sqrt \beta \,(\sigma +1)}(C_2\pm x)\right)\right]^{1/\sigma}&& \text{if} \ \ \beta B^2>\gamma,\\
r(x)&=\left[\frac{2A(\sigma +1)}{(\sigma +2)(\gamma-\beta B^2)}\sin^{2}\left( \frac{\sigma \sqrt{\gamma-\beta B^2}}
{2\sqrt \beta \,(\sigma +1)}(C_2\pm x)\right)\right]^{1/\sigma}&& \text{if} \ \ \beta B^2<\gamma,\\
r(x)&=\left[\frac{A\sigma^2}{2\beta(\sigma +1)(\sigma+2)}\,(C_2\pm x)^2\right]^{1/\sigma}&& \text{if} \ \ \beta B^2=\gamma.
\end{align*}}

\textit{Remark {3}}.
Searching for solutions of system \eqref{eq04} of a more general form $r=r(x)$, $\varphi=At+\theta(y)$ also leads to solution \eqref{new01}.

\textit{Remark {4}}.
System \eqref{eq04} also has solutions of the form $r=r(x)$, $\varphi=At+By+\zeta(x)$.

\subsection{Solutions, periodic in time and in one spatial variable, with amplitude dependent on the other spatial variable. Semi-inverse approach}\label{ss:3.3}

{$1^\circ$. \textit{Simplest version of the semi-inverse approach.}
Now we describe the simple and convenient for analysis semi-inverse approach developed in \cite{polkud2}, which is based on introducing another arbitrary auxiliary function instead of the function $g(r)$ and directly specifying exact solutions in implicit form.}
Namely, we will consider the function $f(r)$ as arbitrary, and the solution $r=r(x)$ of equation \eqref{new02}, taking into account the relation $h=rf$, will be given in terms of an auxiliary function $h=h(x)$ in the implicit form
\begin{equation}
rf(r)=h(x).
\label{eq90}
\end{equation}
The function $g=g(x)$ in this approach is found from equation \eqref{new02}, that leads to the formula
\begin{equation}
g=A+r^{-1}[B^2h(x)-h''_{xx}(x)].
\label{eq91}
\end{equation}
The potential function $g=g(r)$ for a given specific function $h(x)$ is determined by eliminating $x$
from relation \eqref{eq91} using \eqref{eq90}.

{Setting in \eqref{eq90}--\eqref{eq91} various specific functions $h=h(x)$, we will obtain different classes of nonlinear Schr\"odinger equations, depending on one arbitrary function $f(r)$, which admit exact solutions in the implicit form \eqref{eq90}.}

{Note that in many cases it is not necessary to require the auxiliary function $h(x)$ to be monotonic (invertible), since it is not infrequently possible to directly express the second derivative of $h_{xx}''$ in terms of the function $h$ itself as
\begin{equation}
h''_{xx}=Q(h).
\label{eq200}
\end{equation}
Substituting relation \eqref{eq200} into \eqref{eq91} and taking into account the connection $h=rf$, we obtain the following representation for the potential function:
\begin{equation}
g(r)=A+B^2f(r)-r^{-1}Q(rf(r)),
\label{eq201}
\end{equation}
which determines the specific form of the nonlinear PDE \eqref{eq00b}.}

{Under conditions \eqref{eq600} and $h(x) \ge 0$, the implicit relation \eqref{eq90} defines a single-valued function $r = r(x)$.}

{Let us briefly summarize the above: in the semi-inverse approach, the solution $r=r(x)$ of the reduced ODE \eqref{new02} is implicitly prescribed in the form \eqref{eq90}, while the form of the original PDE \eqref{eq00b} is determined by the potential function \eqref{eq201}. It is important to note that when using this method, the dispersion function $f(r)$ in the equation and its solution remains arbitrary.}

Let us demonstrate with {four} specific examples how the semi-inverse approach works in practice.

\textit{Example 1}.
In relations \eqref{eq90}--\eqref{eq91} we substitute the exponentially decreasing function
\begin{equation}
h(x)= \frac 1{p\cosh(kx+q)},
\label{eq08aa}
\end{equation}
where $k$, $p$, and $q$ are free parameters.
Differentiating \eqref{eq08aa}, we find
\begin{equation}
h''_{xx}=\frac{k^2}{p\cosh(kx+q)}-\frac{2k^2}{p\cosh^3(kx+q)}=k^2h-2(kp)^2h^3.
\label{eq08ab}
\end{equation}
From \eqref{eq91} and \eqref{eq08aa}--\eqref{eq08ab} taking into account \eqref{eq90} we obtain
$$
g(r)=a+bf(r)+cr^2f^3(r),\quad \ a=A,\quad b=B^2-k^2,\quad c=2(kp)^2.
$$
It follows from this that the two-dimensional nonlinear Schr\"odinger equation
\begin{equation}
iw_t+\Delta[f(|w|)w]+[a+bf(|w|)+c|w|^2f^3(|w|)]w=0,
\label{eq08ac}
\end{equation}
where $f(r)$ is an arbitrary function, has an exact solution, which is described by the formulas \eqref{eq02} and \eqref{new01}
with $A=a$ and $B=\pm\sqrt{b+k^2}$, and the function $r=r(x)$ is given by the implicit relation
\begin{equation}
\frac {k\sqrt{2/c}}{\cosh(kx+q)}=rf(r),
\label{eq08ad}
\end{equation}
$k$ and $q$ are arbitrary constants\ ($k>0$).

If conditions \eqref{eq600} are satisfied, it follows from \eqref{eq08ad} that $r\to 0$ as $x\to \pm \infty$, i.e. the obtained solution decreases exponentially rapidly for large $x$. Moreover, the function $r=r(x)$ achieves its only maximum value $r_m$ at the point $x=-q/k$; this value is determined from the algebraic (transcendental) equation $k\sqrt{2/c}=r_mf(r_m)$.

{For a power-law dispersion function
\begin{equation}
f(r)=\beta r^\sigma,
\label{eq500}
\end{equation}
where $\beta>0$ and $\sigma\ge 0$ are arbitrary parameters,
equation \eqref{eq08ac} takes the form
$$
iw_t+\beta\Delta(|w|^\sigma w)+(a+b\beta |w|^\sigma+c\beta^3|w|^{3\sigma+2})w=0,
$$
and its exact solution is described by formulas \eqref{eq02} and \eqref{new01}, where the amplitude can be represented in explicit form
$$
r(x)=\BL[\frac {k\sqrt{2/c}}{\beta\cosh(kx+q)}\BR]^{\!\ts\frac 1{\sigma+1}}.
$$}

\textit{Example 2}.
In relations \eqref{eq90}--\eqref{eq91}
we substitute another exponentially decreasing function
\begin{equation}
h(x)= \frac 1{p\cosh^2(kx+q)},
\label{eq08da}
\end{equation}
where $k$, $p$, and $q$ are free parameters.
Differentiating \eqref{eq08da}, we find
\begin{equation}
h''_{xx}=\frac{4k^2}{p\cosh^2(kx+q)}-\frac{6k^2}{p\cosh^4(kx+q)}=4k^2h-6k^2ph^2
\label{eq08db}
\end{equation}
From \eqref{eq91} and \eqref{eq08da}--\eqref{eq08db}, {taking into account \eqref{eq90}, we obtain}
$$
g(r)=a+bf(r)+crf^2(r),\quad \ a=A,\quad b=B^2-4k^2,\quad c=6k^2p.
$$
It follows that of the two-dimensional nonlinear Schr\"odinger equation
\begin{equation}
iw_t+\Delta[f(|w|)w]+[a+bf(|w|)+c|w|f^2(|w|)]w=0,
\label{eq08dc}
\end{equation}
where $f(r)$ is an arbitrary function, has an exact solution, which is described by the formulas \eqref{eq02} and \eqref{new01}
with $A=a$ and $B=\pm\sqrt{b+4k^2}$, and the function $r=r(x)$ is given by the implicit relation
\begin{equation}
\frac {6k^2}{c\cosh^2(kx+q)}=rf(r),
\label{eq08dd}
\end{equation}
$k$ and $q$ are arbitrary constants.

The resulting solution \eqref{eq08dd} has properties qualitatively similar to the properties of solution \eqref{eq08ad} described in Example 1.

{\textit{Example 3}.
Let us now take a more rapidly decreasing function of the form
\begin{equation}
h(x)= k\exp(-\lambda x^2),
\label{eq400}
\end{equation}
where $k$ and $\lambda$ are positive free parameters.
Differentiating \eqref{eq400}, we find
\begin{equation}
h''_{xx}=-2k\lambda\exp(-\lambda x^2)-4k\lambda^2x^2\exp(-\lambda x^2)=-h(4\lambda\ln h+2\lambda-4\lambda\ln k).
\label{eq401}
\end{equation}
Substituting \eqref{eq400} and \eqref{eq401} into \eqref{eq91} and taking into account \eqref{eq90}, we have
$$
g(r)=a+bf(r)+cf(r)\ln[rf(r)],\quad \ a=A,\quad b=B^2+2\lambda-4\lambda\ln k,\quad c=4\lambda.
$$
It follows that of the two-dimensional nonlinear Schr\"odinger equation
\begin{equation*}
iw_t+\Delta[f(|w|)w]+[a+bf(|w|)+cf(|w|)\ln(|w|f(|w|))]w=0,
\end{equation*}
where $f(r)$ is an arbitrary function, has an exact solution described by formulas \eqref{eq02} and \eqref{new01}, in which
the function $r=r(x)$ is given by the implicit relation
\begin{equation}
k\exp(-\lambda x^2)=rf(r).
\label{eq420}
\end{equation}
Solution \eqref{eq420} has properties qualitatively similar to the properties of the solution described in Example 1.}

{For the dispersion function of power-law form \eqref{eq500}, the amplitude from \eqref{eq420} can be expressed explicitly as
$$
r(x)=\BL(\frac k\beta\BR)^{\!\ts\frac 1{\sigma+1}}\exp\BL(-\frac \lambda{\sigma+1}x^2\BR).
$$}

{\textit{Example 4}.
In relations \eqref{eq90}--\eqref{eq91} we substitute the periodic function
\begin{equation}
h(x)=a+b\sin(kx+c),
\label{eq100}
\end{equation}
where $a$, $b$, $c$, and $k$ are free parameters ($a>b>0$).
Taking into account that
\begin{equation}
h''_{xx}=-bk^2\sin(kx+c)=ak^2-k^2h,
\label{eq101}
\end{equation}
and using \eqref{eq91} and \eqref{eq100}--\eqref{eq101} we find the potential
\begin{equation}
g(r)=A-ak^2r^{-1}+(B^2+k^2)f(r).
\label{eq102}
\end{equation}
Thus, we arrive at the nonlinear Schr\"odinger equation
\begin{equation}
iu_t+[f(|u|)u]_{xx}+[A-ak^2|u|^{-1}+(B^2+k^2)f(|u|)]u=0,
\label{eq51}
\end{equation}
which has an exact solution described by the formulas \eqref{eq02} and \eqref{new01}, where the amplitude $r=r(x)$ is given by the implicit relation
$$
a+b\sin(kx+c)=rf(r).
$$}

{$2^\circ$. \textit{More complex version of the approach based on differential constraints.}
This modification of the semi-inverse approach uses solutions defined from suitable auxiliary differential equations \cite{polkud2a}. Namely, instead of an auxiliary function $h(x)$, we first introduce an auxiliary integrable ODE of the form \eqref{eq200}, which is simpler than equation \eqref{new02}. In this approach, the function $f(r)$ is considered arbitrary, whereas the function $Q(r)$ is chosen so that the solution of the ODE \eqref{eq200} can be expressed in terms of elementary or special functions as $h=h(x)$. In this case, the potential function is expressed through $f(r)$ and determined by formula \eqref{eq201}. The solution of the corresponding ODE \eqref{new02} is described by the implicit relation \eqref{eq90}. Note that in the described modification of the semi-inverse approach, the auxiliary ODE \eqref{eq200} can be interpreted as a differential constraint (see, for example, \cite{polzai2012,polzhu2022,pol2026}).}

{Let us use a specific example to demonstrate how modified approach works in practice.}

{\textit{Example 5}. Let us take a differential constraint with quadratic nonlinearity
\begin{equation}
h''_{xx}=ah^2,
\label{eq900}
\end{equation}
which is obtained by substituting the function $Q(h)=ah^2$ into \eqref{eq200}.
The general solution of equation \eqref{eq900} is expressed in terms of the Weierstrass function by the formula (see, for example, \cite{polzai2018}):
\begin{equation*}
h(x)=\frac 6a\wp(x+C_1, 0, C_2),
\end{equation*}
where $C_1$ and $C_2$ are arbitrary constants. In this case, the potential function is defined by the expression
\begin{equation*}
g(r)=A+B^2f(r)-arf^2(r),
\end{equation*}
where $f(r)$ is an arbitrary function specifying the dispersion.
The exact solution of the ODE for the amplitude \eqref{new02} admits a representation in implicit form
$$
\frac 6a\wp(x+C_1, 0, C_2)=rf(r).
$$}

\subsection{Time-periodic solutions with an amplitude depending on two spatial variables}\label{ss:3.4}

The PDE system \eqref{eq04} admits a non-stationary time-periodic solution \eqref{eq02} in the form of a product of functions of different arguments
\begin{equation}
r=r(x,y),\quad \ \varphi=C_1t+C_2,
\label{eq05*}
\end{equation}
where $C_1$ and $C_2$ are arbitrary constants, and the function $r=r(x,y)$ is described by a two-dimensional stationary partial differential equation
\begin{equation}
\Delta h-C_1r+rg(r)=0,\quad \ h=rf(r).
\label{eq05**}
\end{equation}
Exact solutions of this equation for some functions $f(r)$ and $g(r)$ can be found, for example, using the results of \cite{polzai2012}.

\textit{Example {6}}.
Let us consider {in more detail a special case by establishing a linear dependence between the functions $f=f(r)$ and $g=g(r)$}:
$$
g=af+b,
$$
{where $a$ and $b$ are free parameters.}
In this case, the two-dimensional nonlinear Schr\"odinger equation \eqref{eq00b} takes the form
\begin{equation}
iw_t+\Delta[f(|w|)w]+[af(|w|)+b]w=0,
\label{eq94**}
\end{equation}
where $f(z)$ is an arbitrary function. Assuming $g=af+b$ and $C_1=b$ in \eqref{eq05**}, and also taking into account the relation $h=rf$, we arrive at the two-dimensional Helmholtz equation for the function $h$:
\begin{equation}
\Delta h+ah=0,\quad \ \Delta h=h_{xx}+h_{yy}.
\label{eq95**}
\end{equation}

Solutions to this linear PDE can be found, for example, in \cite{polnaz2012}. Any particular solution $h=h(x,y)$ of the equation \eqref{eq95**} generates an exact solution to the nonlinear PDE \eqref{eq94**}, which is described by the formulas \eqref{eq02} and \eqref{eq05*} with $C_1=b$, where the function $r=r(x,y)$ is given implicitly by the relation $h(x,y)=rf(r)$. {If conditions \eqref{eq600} hold, then this implicit relation defines a single-valued function $r=r(x,y)$.}

In particular, equation \eqref{eq95**} admits simple exact solutions
\begin{align*}
h&=(A_1\cos \mu_1x+B_1\sin\mu_1x)(A_2\cos \mu_2y+B_2\sin\mu_2y), &&a=\mu_1^2+\mu_2^2;\\
h&=(A_1\cos \mu_1x+B_1\sin\mu_1x)(A_2\cosh \mu_2y+B_2\sinh\mu_2y), &&a=\mu_1^2-\mu_2^2;\\
h&=(A_1\cosh \mu_1x+B_1\sinh\mu_1x)(A_2\cosh \mu_2y+B_2\sinh\mu_2y), &&a=-\mu_1^2-\mu_2^2;
\end{align*}
where $A_1$, $A_2$, $B_1$, $B_2$, $\mu_1$, and $\mu_2$ are arbitrary constants.

{Equation \eqref{eq95**} also has the radially symmetric solution
$$
h=\begin{cases}AJ_0(\sqrt{a}\,\rho)+BY_0(\sqrt{a}\,\rho)& \text{if \ $a>0$},\\
AI_0(\sqrt{|a|}\,\rho)+BK_0(\sqrt{|a|}\,\rho)& \text{if \ $a<0$},\end{cases}
\quad \ \ \rho=\sqrt{x^2+y^2},
$$
where $A$ and $B$ \arbs, $J_0(z)$ and $Y_0(z)$ are the Bessel functions, and $I_0(z)$ and $K_0(z)$ are the modified Bessel functions.}

{More complex two-dimensional exact solutions of the linear PDE \eqref{eq95**} in polar coordinates ($\rho$, $\theta$) have the form:
$$
h=\begin{cases}[A_nJ_n(\mu \rho )+B_nY_n(\mu \rho )][C_n\cos(n\theta)+D_n\sin(n\theta)]& \text{if \ $a=\mu^2>0$},\\
[A_nI_n(\mu \rho )+B_nK_n(\mu \rho )][C_n\cos(n\theta)+D_n\sin(n\theta)]& \text{if \ $a=-\mu^2<0$},\end{cases}
$$
where $n=1, \, 2, \, \ldots\,$; \ $A_n$, $B_n$, $C_n$, and $D_n$ \arbs;
$J_n(z)$ and $Y_n(z)$ are the Bessel functions;
and $I_n(z)$ and $K_n(z)$ are the modified Bessel functions.}

\textit{Example {7}}. We are looking for solutions of equation \eqref{eq05**} of a traveling wave type with respect to the spatial variables
\begin{equation}
r=r(z),\quad \ z=Ax+By,
\label{nn01}
\end{equation}
where $A$ and $B$ are arbitrary constants.
As a result, we arrive at the autonomous second-order ODE
\begin{equation}
(A^2+B^2)h''_{zz}-C_1r+rg(r)=0,\quad \ h=rf(r),
\label{nn02}
\end{equation}
general solution of which can be represented in implicit form.

Next, we reason in the same way as we did earlier in Section~\ref{ss:3.2}. We use the function \eqref{eq08aa} and its derivative \eqref{eq08ab}, in which $x$ is replaced by $z$. As a result, we arrive at the two-dimensional nonlinear Schr\"odinger equation
\begin{align*}
&iw_t+\Delta[f(|w|)w]+[a-bf(|w|)+c|w|^2f^3(|w|)]w=0,\\
&a=C_1,\quad b=k^2(A^2+B^2),\quad c=2(kp)^2(A^2+B^2),
\end{align*}
which includes an arbitrary function $f(r)$. An exact solution of this PDE is described by the formulas \eqref{eq02}, \eqref{eq05*}, and \eqref{nn01},
where the function $r=r(z)$  is given by the implicit relation
\begin{equation}
\frac 1{p\cosh(kz+q)}=rf(r).
\label{e100}
\end{equation}
Here $k$, $p$, and $q$ are free parameters.
For $p>0$, $f>0$, and $f'_r\ge 0$ it follows from \eqref{e100} that $r\to 0$ as $z\to \pm \infty$, i.e. the resulting solution decreases exponentially rapidly for large $z$. This solution has a single maximum and describes a solitary wave.

\textit{Example {8}}.
In a similar way, using the function \eqref{eq08da} and its derivative \eqref{eq08db}, in which $x$ is replaced by $z$, we obtain the two-dimensional nonlinear Schr\"odinger equation
\begin{align*}
&iw_t+\Delta[f(|w|)w]+[a-bf(|w|)+c|w|f^2(|w|)]w=0,\\
&a=C_1,\quad b=4k^2(A^2+B^2),\quad c=6k^2p(A^2+B^2),
\end{align*}
which contains an arbitrary function $f(r)$. An exact solution of this PDE is described by the formulas \eqref{eq02}, \eqref{eq05*}, and \eqref{nn01},
where the function $r=r(z)$  is given by the implicit relation
\begin{equation*}
\frac 1{p\cosh^2(kz+q)}=rf(r).
\end{equation*}

\textit{Remark {5}}.
System \eqref{eq04} has a wider class, than \eqref{eq05*}, of time-periodic two-dimensional solutions of the form
\begin{equation}
r=r(x,y),\quad \ \varphi=At+\xi(x,y).
\label{new05}
\end{equation}
Substituting these expressions into \eqref{eq04}, we obtain a stationary nonlinear PDE system for the functions $r$ and $\xi$.

\subsection{Generalized separable solutions with time-dependent amplitude}

In the general case, system \eqref{eq04} admits solutions whose amplitude depends only on time, and the phase is a quadratic polynomial in spatial coordinates with variable coefficients:
\begin{equation}
r=r(t),\quad \ \varphi=a_{11}(t)x^2+a_{12}(t)xy+a_{22}(t)y^2+b_1(t)x+b_2(t)y+c(t),
\label{eqq02}
\end{equation}
where the seven functions $r=r(t)$, $a_{11}=a_{11}(t)$, $a_{12}=a_{12}(t)$, $a_{22}=a_{22}(t)$, $b_1=b_1(t)$, $b_2=b_2(t)$, and $c=c(t)$
are to be determined in the course of further analysis.
Substituting \eqref{eqq02} into \eqref{eq04}, after separation of variables we arrive at the following nonlinear ODE system for functional coefficients:
\begin{equation}
\begin{aligned}
a_{11}'&=-(4a_{11}^2+a_{12}^2)f(r),\\
a_{12}'&=-4a_{12}(a_{11}+a_{22})f(r),\\
a_{22}'&=-(4a_{22}^2+a_{12}^2)f(r),\\
b_1'&=-2(2a_{11}b_1+a_{12}b_2)f(r),\\
b_2'&=-2(2a_{22}b_2+a_{12}b_1)f(r),\\
c'&=-(b_1^2+b_2^2)f(r)+g(r),\\
r'&=-2(a_{11}+a_{22})rf(r),
\end{aligned}
\label{eqq03}
\end{equation}
where the first six equations were divided by $r$, and the prime denotes the derivative with respect to $t$.

Below we describe four multiparameter exact solutions of the ODE system \eqref{eqq03}, which are expressed in terms of elementary functions and/or quadratures and determine the corresponding exact solutions of the nonlinear Schr\"odinger equations of general form \eqref{eq00a} and \eqref{eq00b}.

$1^\circ$. \textit{Solution for $a_{12}=0$ in case $f(r)\equiv 1$}.
When $a_{12}=0$ the second equation of the ODE system \eqref{eqq03} is satisfied identically. In this case, the remaining first four equations are first integrated sequentially, then the last equation is integrated, and finally the penultimate one. As a result of these actions, we obtain
\begin{equation}
\begin{gathered}
a_{11}(t)=\frac 1{4(t+C_1)},\quad a_{12}(t)=0,\quad a_{22}(t)=\frac 1{4(t+C_2)},\\
b_1(t)=\frac {C_3}{2(t+C_1)},\quad b_2(t)=\frac {C_4}{2(t+C_2)},\quad
r(t)=\frac{C_5}{\sqrt{(t+C_1)(t+C_2)}},\\
c(t)=\frac {C_3^2}{4(t+C_1)}+\frac {C_4^2}{4(t+C_2)}+\int g\bl(r(t)\br)\,dt+C_6,
\end{gathered}
\label{eqq03*}
\end{equation}
where $C_1$, \dots, $C_6$ are arbitrary constants.

$2^\circ$. \textit{Solution for $a_{12}=0$ in the case of arbitrary $f(r)$}.
Let us show that the solution given in Item~$1^\circ$ can be generalized to the case of an arbitrary function $f(r)$. Indeed, for $a_{12}=0$, as before, the second equation of the ODE system \eqref{eqq03} is satisfied identically. From the first, third, fourth, fifth and last equations of the system \eqref{eqq03}, by eliminating the function $f(r)$, four integrals are found
\begin{equation}
\frac 1{a_{11}}=\frac 1{a_{22}}+C_1,\quad \ b_1=C_2a_{11},\quad \ b_2=C_3a_{22},\quad \
a_{11}a_{22}=C_4r^2,
\label{ad01}
\end{equation}
where $C_1$, $C_2$, $C_3$, and $C_4$ are arbitrary constants.
From the first and last integrals \eqref{ad01} we express through $r$ the functional coefficients $a_{11}=a_{11}(r)$ and $a_{22}=a_{22}(r)$, and then substitute them into the last equation \eqref{eqq03}. As a result, we obtain a first-order ODE with separable variables. Having solved it, we find the dependence $r=r(t)$. After this, the penultimate ODE \eqref{eqq03} is integrated, in which $b_k$ should be expressed through $a_{kk}$ using the relations \eqref{ad01}.

$3^\circ$. \textit{Solution for $a_{12}\not=0$ in case $f(r)\equiv 1$}.
Assuming that $a_{12}\not=0$, we seek solutions of the first three functional coefficients in the ODE system \eqref{eqq03} in the form of some constants divided by $(t+C_1)$. In this way, we can find the following exact solution:
\begin{equation}
\begin{gathered}
a_{11}(t)=\frac {C_2}{t+C_1},\quad a_{12}(t)=\frac A{t+C_1},\quad a_{22}(t)=\frac {B}{t+C_1},\\
b_1(t)=\frac{2AC_3}{t+C_1}+AC_4,\quad b_2(t)=\frac{(1-4C_2)C_3}{t+C_1}-2C_2C_4,\\
r(t)=\frac{C_5}{\sqrt{t+C_1}},\quad A=\pm(C_2-4C_2^2)^{1/2},\quad B=\frac 1{4}-C_2,\\
c(t)=-\int\bl[b_1^2(t)+b_2^2(t)\br]\,dt+\int g\bl(r(t)\br)\,dt+C_6,
\end{gathered}
\label{eqq04}
\end{equation}
where $C_1$, \dots, $C_6$ are arbitrary constants. The first integral in the last expression \eqref{eqq04} is easy to calculate, but is not given due to its bulkiness.

$4^\circ$. \textit{Solution for $a_{12}\not=0$ in the case of arbitrary $f(r)$}.
The results presented in Item~$3^\circ$ can be generalized to the case of an arbitrary function $f(r)$. For this, we seek solutions to the ODE system \eqref{eqq03} in the form
\begin{equation}
\begin{gathered}
a_{11}=C_2\varphi(t),\quad a_{12}=A\varphi(t),\quad a_{22}=B\varphi(t),\\
b_1=2AC_3\varphi(t)+AC_4,\quad b_2=(1-4C_2)C_3\varphi(t)-2C_2C_4,\\
A=\pm(C_2-4C_2^2)^{1/2},\quad B=\tfrac 1{4}-C_2,
\end{gathered}
\label{ad02}
\end{equation}
where $C_2$, $C_3$, and $C_4$ are arbitrary constants, and the function $\varphi=\varphi(t)$ satisfies the equation
\begin{equation}
\varphi'_t=-\varphi^2f(r).
\label{ad03}
\end{equation}
It is not difficult to verify that when conditions \eqref{ad02} and \eqref{ad03} are satisfied, the first five equations of the ODE system \eqref{eqq03} are satisfied identically. Substituting the expressions for $a_{11}$ and $a_{22}$ from \eqref{ad02} into the last ODE \eqref{eqq03}, we have
\begin{equation}
r'_t=-\tfrac12\varphi rf(r).
\label{ad04}
\end{equation}
From the equations \eqref{ad03} and \eqref{ad04} the integral follows
\begin{equation}
\varphi=2C_5r^2.
\label{ad05}
\end{equation}
By eliminating the function $\varphi$ from \eqref{ad04} using \eqref{ad05}, we arrive at the first-order ODE with separable variables
\begin{equation*}
r'_t=-C_5r^3f(r),
\end{equation*}
the general solution of which $r=r(t)$ can be represented in implicit form as
\begin{equation}
\int\frac{dr}{r^3f(r)}=C_6-C_5t.
\label{ad06}
\end{equation}
The function $\varphi(t)$ included in the functional coefficients \eqref{ad02} is found by the formula \eqref{ad05}, and the function $c=c(t)$ is determined by direct integration of the penultimate ODE \eqref{eqq03}.

\subsection{Solutions with functional separation of variables of equation (\ref{eq00a})}

The two-dimensional nonlinear Schr\"odinger equation \eqref{eq00a} admits an exact solution \eqref{eq02}, in which the amplitude and phase are sought in the form
\begin{equation}
r=r(z),\quad \ z=t^2+a_1x+a_2y,\quad \ \varphi=ct^3+t(b_1x+b_2y)+d,
\label{x02}
\end{equation}
where $d$ is an arbitrary constant, and the free parameters $a_1$, $a_2$, $b_1$, $b_2$, and $c$ are selected so as to satisfy the PDE system \eqref{eq04}.

Substituting \eqref{x02} into the second equation of system \eqref{eq04}, after reducing by $2tu'_z$ we obtain the simple algebraic relation
\begin{equation}
a_1b_1+a_2b_2+1=0.
\label{x03}
\end{equation}
Substituting \eqref{x02} into the first equation of system \eqref{eq04}, after elementary algebraic transformations we have
\begin{equation}
(a_1^2+a_2^2)r''_{zz}-[(3c+b_1^2+b_2^2)t^2+b_1x+b_2y]r+rg(r)=0.
\label{x04}
\end{equation}
This relation will be an ordinary differential equation if we require that the expression in square brackets be a function only of the variable $z$, which is introduced in \eqref{x02}. This requirement leads to the following three algebraic equations on the free coefficients:
\begin{equation}
\begin{aligned}
3c+b_1^2+b_2^2&=\lambda,\\
b_1&=\lambda a_1,\\
b_2&=\lambda a_2,
\end{aligned}
\label{x05}
\end{equation}
where $\lambda$ is a new free parameter. When conditions \eqref{x05} are satisfied, the relation \eqref{x04} becomes the ordinary differential equation
\begin{equation}
(a_1^2+a_2^2)r''_{zz}-\lambda zr+rg(r)=0.
\label{x06}
\end{equation}

The four relations \eqref{eq03} and \eqref{x05} represent an underdetermined system of algebraic equations, which includes six unknown coefficients $a_1$, $b_1$, $a_2$, $b_2$, $c$, and $\lambda$. Therefore, two of these coefficients can be specified arbitrarily, and the other four can be expressed through them. It is easy to verify by direct verification that the general solution of the system of equations \eqref{eq03} and \eqref{x05} can be represented as
\begin{equation}
\begin{gathered}
a_1=p\cos q,\quad \ a_2=p\sin q,\quad \
b_1=-\frac 1p{\cos q},\quad \ b_2=-\frac 1p{\sin q},\\
c=-\frac 2{3p^2},\quad \ \lambda= -\frac 1{p^2},
\end{gathered}
\label{x07}
\end{equation}
where $p$ and $q$ are arbitrary constants ($p\not=0$). Substituting the expressions \eqref{x07} into \eqref{x02} and \eqref{x06}, and then into \eqref{eq02}, we can find the exact solution of the two-dimensional nonlinear Schr\"odinger equation \eqref{eq00a}.

\subsection{Reduction using a variable of type two-dimensional traveling wave}

The PDE system \eqref{eq04} admits a solution of the form
\begin{equation}
\begin{aligned}
r=r(z),\quad \ \varphi=at+b_1x+b_2y+\theta(z),\quad \ z=c_1x+c_2y+\lambda t,
\label{new10}
\end{aligned}
\end{equation}
where $a$, $b_1$, $b_2$, $c_1$, $c_2$, and $\lambda$ are arbitrary constants, and the functions $r=r(z)$ and $\theta=\theta(z)$ satisfy the nonlinear ODE system:
\begin{equation}
\begin{aligned}
&-r(a+\lambda\theta'_z)+(c_1^2+c_2^2)h''_{zz}-h[(b_1+c_1\theta'_z)^2+(b_2+c_2\theta'_z)^2]+rg(r)=0,\\
&\qquad\lambda r'_z+2(b_1c_1+b_2c_2)h'_z+(c_1^2+c_2^2)(2h'_z\theta'_z+h\theta''_{zz})=0,\quad \ h=rf(r).
\label{new11}
\end{aligned}
\end{equation}
Note that the replacement $\xi=\theta'_z$ allows us to reduce the order of the ODE system \eqref{new11} by one.

It is easy to check that the second equation \eqref{new11} admits the first integral
\begin{equation}
\lambda\int h(r)\,dr+(b_1c_1+b_2c_2)h^2+(c_1^2+c_2^2)h^2\theta'_z=C,\quad \ h=rf(r),
\label{new12}
\end{equation}
where $C$ is an arbitrary constant.
By eliminating the derivative $\theta'_z$ from the first equation \eqref{new11} using \eqref{new12}, we can derive a nonlinear autonomous second-order ODE (which is explicitly independent of $z$). The general solution of the resulting ODE can be expressed in quadratures using the method described in \cite{polkud2}.

\section{Solutions and reductions of the nonlinear Schr\"odinger equation in the polar coordinate system}

\subsection{Transformation of the nonlinear Schr\"odinger equation to a system of real PDEs in polar coordinates}

We now use the polar coordinate system, which is introduced using the formulas
$$
x=\rho\cos\theta,\quad y=\rho\sin\theta\quad (0\le \rho<\infty,\quad 0\le \theta\le 2\pi).
$$
As before, we represent the desired function in exponential form \eqref{eq02},
where $r=r(\rho,\theta,t)\ge 0$ and $\varphi=\varphi(\rho,\theta,t)$ are real functions.

In this case, the differential operators included in \eqref{eq00b} and \eqref{eq03} are written as follows:
\begin{equation*}
\Delta h=\rho^{-1}(\rho h_\rho)_\rho+\rho^{-2}h_{\theta\theta},\quad \
|\nabla\varphi|^2=\varphi_\rho^2+\rho^{-2}\varphi_\theta^2,\quad \ \nabla h\cdot\nabla\varphi=h_\rho\varphi_\rho+\rho^{-2}h_\theta\varphi_\theta,
\end{equation*}
and the system of real PDEs \eqref{eq04} is transformed to the form
\begin{equation}
\begin{aligned}
-r\varphi_t+\rho^{-1}(\rho h_\rho)_\rho+\rho^{-2}h_{\theta\theta}-h(\varphi_\rho^2+\rho^{-2}\varphi_\theta^2)+rg(r)&=0,\\
r_t+2(h_\rho\varphi_\rho+\rho^{-2}h_\theta\varphi_\theta)+h[\rho^{-1}(\rho \varphi_\rho)_\rho+\rho^{-2}\varphi_{\theta\theta}]&=0,\quad \ h=rf(r).
\end{aligned}
\label{eq04ab}
\end{equation}

The solution representation \eqref{eq02} and the PDE system \eqref{eq04ab} will also be used to construct exact solutions of the nonlinear Schr\"odinger equations \eqref{eq00a} and \eqref{eq00b}.

\subsection{Nonlinear Schr\"odinger equations describing radially symmetric solutions}

Equations \eqref{eq00a} and \eqref{eq00b} are significantly simplified if we consider radially symmetric solutions. In this case, we arrive at PDEs with two independent variables
\begin{align}
iw_t+\rho^{-1}(\rho w_\rho)_\rho+g(|w|)w&=0,\label{eq01b}\\
iw_t+\rho^{-1}\{\rho[f(|w|)w]_\rho\}_\rho+g(|w|)w&=0,\label{eq01c}
\end{align}
where $\rho=\sqrt{x^2+y^2}$ is the radial coordinate.

As before, we seek the desired function in exponential form \eqref{eq02}. As a result, from equation \eqref{eq01c} we obtain a system of two real PDEs \eqref{eq04ab}, in which the desired functions $r$ and $\varphi$ do not depend on the angular coordinate $\theta$:
\begin{equation}
\begin{aligned}
-r\varphi_t+\rho^{-1}(\rho h_\rho)_\rho-h\varphi_\rho^2+rg(r)&=0,\\
r_t+2h_\rho\varphi_\rho+h\rho^{-1}(\rho \varphi_\rho)_\rho&=0,\quad \ h=rf(r).
\end{aligned}
\label{eq04b}
\end{equation}

Below in Sections \ref{ss:4.3}--\ref{ss:4.5} some exact solutions of the system \eqref{eq04b} are described.

\subsection{Radially symmetric time-periodic solutions whose phase does not depend on the radial coordinate}\label{ss:4.3}

System \eqref{eq04b} admits simple time-periodic exact solutions of the form
\begin{equation}
r=r(\rho),\quad \ \varphi=C_1t+C_2,
\label{e01}
\end{equation}
where $C_1$ and $C_2$ are arbitrary constants, and the function $r=r(\rho)$ is described by the second-order nonlinear ODE
\begin{equation}
\rho^{-1}(\rho h'_\rho)'_\rho-C_1r+rg(r)=0,\quad \ h=rf(r).
\label{e02}
\end{equation}
The general solution of this non-autonomous equation cannot be represented in closed form in the case of arbitrary functions $f(r)$ and $g(r)$.

Let us find some exact solutions of ODE \eqref{e02} using the semi-inverse approach described in Section \ref{ss:3.2}, based on introducing an arbitrary auxiliary function instead of the function $g(r)$ and directly specifying the exact solutions in implicit form. Namely, we will consider the function $f(r)$ to be arbitrary, and the solution $r=r(\rho)$ of equation \eqref{e02}, taking into account the relation $h=rf$, will be specified using an arbitrary auxiliary function $h=h(\rho)$ in implicit form
\begin{equation}
rf(r)=h(\rho).
\label{eq90*}
\end{equation}
The function $g=g(r)$ in this approach is no longer specified, but is found directly from equation \eqref{e02}, which leads to the formula
\begin{equation}
g=C_1+\rho^{-1}(\rho h'_\rho)'_\rho\, r^{-1}.
\label{eq91*}
\end{equation}
The potential function $g=g(r)$ for a given specific function $h(\rho)$ is determined by eliminating $\rho$ from relations \eqref{eq90*} and \eqref{eq91*}.

Let us now demonstrate with a specific example how the described approach works in practice. To do this, we take a simple elementary function $h=h(\rho)$ and find the function $g=g(r)$ generated by it. In this case, the function $g(r)$ will be expressed through $f(r)$.

\textit{Example {9}.}
In the relations \eqref{eq90*} and \eqref{eq91*} we substitute the increasing function
\begin{equation}
h(\rho)=a\rho^{2}+b,
\label{eq92xx}
\end{equation}
where $a$ and $b$ are arbitrary constants.
As a result we get
\begin{equation}
rf(r)=a\rho^{2}+b,\quad \ g(r)=C_1+4ar^{-1}.
\label{eq93xx}
\end{equation}
Here the first relation, where $f(r)$ is an arbitrary function, implicitly specifies the dependence of the amplitude on the spatial coordinate $r=r(\rho)$.

Formulas \eqref{eq02}, \eqref{e01} and the first relation \eqref{eq93xx} define the exact solution in implicit form of equation \eqref{eq01c} with an arbitrary function $f(|w|)$ and potential $g(|w|)=C_1+4a|w|^{-1}$.
\medskip

\textit{Example {10}.}
In the relations \eqref{eq90*} and \eqref{eq91*} we now substitute the decreasing function
\begin{equation}
h(\rho)=\frac 1{a\rho^{2}+b},
\label{eq92}
\end{equation}
where $a$ and $b$ are arbitrary constants.
Simple calculations lead to the relations
\begin{equation}
rf(r)=\frac 1{a\rho^{2}+b},\quad \ g(r)=C_1+4arf^2(r)-8abr^2f^3(r).
\label{eq93}
\end{equation}
Here the first equality, where $f(r)$ is an arbitrary function, implicitly specifies the dependence of the amplitude on the spatial coordinate $r=r(\rho)$.

Formulas \eqref{eq02}, \eqref{e01} and the first relation \eqref{eq93} define the exact solution in implicit form of the equation \eqref{eq01c} with an arbitrary function $f(|w|)$ and potential $g(|w|)=C_1+4a|w|f^2(|w|)-8ab|w|^2f^3(|w|)$.
\medskip

It is also possible to appropriately give the relationship between the functions $f(r)$ and $g(r)$, and find the function $h=h(\rho)$ from the resulting differential equation.

\textit{Example {11}.}
Assuming in \eqref{eq01c} and \eqref{e02} (or \eqref{eq91*}) $g=af+b$, we arrive at the nonlinear Schr\"odinger equation
\begin{equation}
iw_t+\rho^{-1}\{\rho[f(|w|)w]_\rho\}_\rho+[af(|w|)+b]w=0,
\label{eq94}
\end{equation}
where $f(z)$ is an arbitrary function, which admits an exact solution determined by the formulas \eqref{eq02} and \eqref{e01} for $C_1=b$.
In these formulas, the amplitude $r=r(\rho)$ is given by an implicit relation \eqref{eq90*}, in which the function $h=h(\rho)$ is described by the linear ODE:
\begin{equation}
h_{\rho\rho}''+\rho^{-1}h_\rho'+ah=0.
\label{eq95}
\end{equation}
The general solution of this equation is expressed in terms of Bessel functions or modified Bessel functions \cite{polzai2018}:
\begin{align*}
h(\rho)=\begin{cases} A_1J_0(\sqrt a\,\rho)+A_2Y_0(\sqrt a\,\rho)& \text{if} \ \ a>0,\\
A_1I_0(\sqrt{|a|}\,\rho)+A_2K_0(\sqrt{|a|}\,\rho)& \text{if} \ \ a<0,\end{cases}
\end{align*}
where $A_1$ and $A_2$ are arbitrary constants.

\subsection{Time-periodic solutions whose amplitude and phase depend on the radial coordinate}

The system \eqref{eq04b} admits a more complex, than \eqref{e01}, exact solution of the form
\begin{equation}
r=r(\rho),\quad \ \varphi=C_1t+\zeta(\rho),
\label{eq05b}
\end{equation}
where $C_1$ is an arbitrary constant, and the functions $r=r(\rho)$ and $\zeta=\zeta(\rho)$ are described by the ODE system
\begin{equation}
\begin{aligned}
\rho^{-1}(\rho h'_\rho)'_\rho-h(\zeta'_\rho)^2-C_1r+rg(r)&=0,\\
2h'_\rho\zeta'_\rho+h\rho^{-1}(\rho\zeta'_\rho)'_\rho&=0,\quad \ h=rf(r).
\end{aligned}
\label{eq06b}
\end{equation}
Integrating the second equation \eqref{eq06b}  twice, we successively obtain
\begin{equation}
\zeta'_\rho=C_2\rho^{-1}h^{-2},\quad \
\zeta=C_2\int \rho^{-1}h^{-2}d\rho+C_3,
\label{eq07b}
\end{equation}
where $C_2$, $C_3$ are arbitrary constants. Eliminating the derivative $\zeta'_\rho$ from the first equation \eqref{eq06b} using the first relation \eqref{eq07b}, we obtain the following second-order nonlinear non-autonomous ODE for the function $r=r(\rho)$:
\begin{equation}
\rho^{-1}(\rho h'_\rho)'_\rho-C_2^2\rho^{-2}h^{-3}-C_1r+rg(r)=0,\quad \ h=rf(r).
\label{eq08b}
\end{equation}
This equation differs from equation \eqref{e02} by the presence of an additional nonlinear term proportional to $h^{-3}$. Some exact solutions of this ODE can be obtained using the semi-inverse approach described earlier, specifying the auxiliary function $h=h(\rho)$.

\textit{Remark {6}}. Equation \eqref{eq01c} in some cases (when two functions $f$ and $g$ are defined by one suitable auxiliary arbitrary function) can also admit more complex, than \eqref{eq05b}, solutions of the form
\begin{equation*}
r=r(\rho),\quad \ \varphi=a(t)\zeta(\rho)+b(t).
\end{equation*}
For the nonlinear Schr\"odinger equation with one spatial variable, such solutions were constructed in \cite{polkud2}.

\subsection{Radially symmetric solutions whose amplitude depends on time}\label{ss:4.5}

Let us show that system \eqref{eq04b} admits an exact solution with a generalized separation of variables of the form
\begin{equation}
r=r(t),\quad \ \varphi=a(t)\rho^2+b(t).
\label{eq09b}
\end{equation}
To do this, we substitute \eqref{eq09b} into \eqref{eq04b}. As a result, the first equation of the system is reduced to a quadratic equation with respect to $\rho$, the coefficients of which depend on time. By equating the functional coefficients of the quadratic equation to zero and adding the second equation of the system, which in this case depends only on $t$, we obtain the following ODE system:
\begin{equation}
\begin{aligned}
a'_t&=-4a^2f(r),\\
b'_t&=g(r),\\
r'_t&=-4arf(r),
\end{aligned}
\label{eq10b}
\end{equation}
in which the first two equations were reduced by $r$.

From the first and third equations of system \eqref{eq10b} we have the integral
\begin{equation}
a=C_1r,
\label{eq10c}
\end{equation}
where $C_1$ is an arbitrary constant. Eliminating $a$ from the third equation \eqref{eq10b} using \eqref{eq10c}, we obtain the ODE
\begin{equation}
r'_t=-4C_1r^2f(r),
\label{eq10d}
\end{equation}
whose general solution can be represented in implicit form
\begin{equation}
\int\frac{dr}{r^2f(r)}=C_2-4C_1t,
\label{eq10e}
\end{equation}
where $C_2$ is an arbitrary constant. The function $a=a(t)$ is defined by the relations \eqref{eq10c} and \eqref{eq10e}.
The function $b=b(t)$ is found by integrating the second ODE \eqref{eq10b}:
\begin{equation}
b=\int g(r)\,dt+C_3,
\label{eq10f}
\end{equation}
where $C_3$ is an arbitrary constant, and the function $r=r(t)$ is defined implicitly by the expression \eqref{eq10e}.

Note that the function $b$ can be expressed through the function $r$ using the formula
$$
b=-\frac1{4C_1}\int\frac{g(r)\,dr}{r^2f(r)}+C_3.
$$

\subsection{Solutions that are periodic in time and angular coordinate and are based on the use of a traveling wave variable}\label{ss:4.6}

The PDE system \eqref{eq04ab} admits a periodic in time and angular coordinate exact solution of the form

\noindent
\begin{equation}
r=r(\rho),\quad \varphi=C_1t+C_2\theta+C_3,
\label{nn20}
\end{equation}
where $C_1$, $C_2$, and $C_3$ are arbitrary constants, and the function $r=r(\rho)$ is described by the second-order nonlinear non-autonomous ODE
\begin{equation}
\rho^{-1}(\rho h'_\rho)'_\rho-C_2^2\rho^{-2}h-C_1r+rg(r)=0,\quad \ h=rf(r).
\label{nn21}
\end{equation}
This equation differs from equation \eqref{e02} by the presence of an additional term proportional to $\rho^{-2}h$.
The general solution of ODE \eqref{nn21} cannot be expressed in quadratures in the case of arbitrary functions $f(r)$ and $g(r)$.
Some exact solutions of ODE \eqref{nn21} can be found using the semi-inverse approach described in Section~\ref{ss:3.2}.
It is also possible to appropriately give the relationship between the functions $f(r)$ and $g(r)$, and find the function $h=h(\rho)$ from the resulting differential equation.

\textit{Example {12}.}
Assuming in \eqref{eq00b} and \eqref{nn21} $g=af+b$, we arrive at the nonlinear Schr\"odinger equation
\begin{equation*}
iw_t+\Delta[f(|w|)w]+[af(|w|)+b]w=0,
\end{equation*}
where $\Delta h=\rho^{-1}(\rho h_\rho)_\rho+\rho^{-2}h_{\theta\theta}$ and $f(z)$ is an arbitrary function, which admits an exact solution determined by the formulas \eqref{eq02} and \eqref{nn20} for $C_1=b$. In these formulas, the amplitude $r=r(\rho)$ is given by the implicit relation \eqref{eq90*}, in which the function $h=h(\rho)$ is described by the linear ODE:
\begin{equation*}
\rho^2h_{\rho\rho}''+\rho h_\rho'+(a\rho^2-\nu^2)h=0,\quad \ \nu=C_2.
\end{equation*}
The general solution of this equation is expressed in terms of Bessel functions or modified Bessel functions \cite{polzai2018}:
\begin{align*}
h(\rho)=\begin{cases} A_1J_{\nu}(\sqrt a\,\rho)+A_2Y_{\nu}(\sqrt a\,\rho)& \text{if} \ \ a>0,\\
A_1I_{\nu}(\sqrt{|a|}\,\rho)+A_2K_{\nu}(\sqrt{|a|}\,\rho)& \text{if} \ \ a<0,\end{cases}
\end{align*}
where $A_1$ and $A_2$ are arbitrary constants, and $\nu=C_2$.

\textit{Remark {7}}.
If the solutions described by the formulas \eqref{eq02} and \eqref{nn20} are considered on the entire plane $(\rho,\theta)$, then it is additionally necessary to set the condition of periodicity of these solutions with respect to the angular variable:
\begin{equation}
w(\rho,\theta,t)=w(\rho,\theta+2\pi,t).
\label{nn22}
\end{equation}
In this case, the constant $C_2$ in \eqref{nn20} cannot be arbitrary; however, it can take a series of discrete values
$C_2=\pm n$, where $n=0,\,1,\,\dots$
These solutions will be invariant under rotation by an angle $2\pi/n$ (if $n\ge 1$).

If the solutions described by formulas \eqref{eq02} and \eqref{nn20} are considered in some sector $0\le\theta\le\theta_0<2\pi$, then the constant $C_2$ can be chosen arbitrarily.

\textit{Remark {8}}.
The PDE system \eqref{eq04ab} has a wider class of solutions, then \eqref{nn20}, of the form
\begin{equation}
r=r(\rho),\quad \varphi=C_1t+C_2\theta+\xi(\rho).
\label{nn23}
\end{equation}
which are periodic in time and angular coordinate.
Substituting expressions \eqref{nn23} into \eqref{eq04ab}, we obtain a nonlinear ODE system, which can be reduced to a single ODE for the function $r=r(\rho)$:
\begin{equation}
\rho^{-1}(\rho h'_\rho)'_\rho-C_3^2\rho^{-2}h^{-3}-C_2^2\rho^{-2}h-C_1r+rg(r)=0,\quad \ h=rf(r),
\label{nn24}
\end{equation}
and the function $\xi=\xi(\rho)$ is expressed through $r=r(\rho)$ using quadrature
$$
\xi=C_3\int\frac{d\rho}{\rho r^2f^2(r)}+C_4,
$$
where $C_3$ and $C_4$ are arbitrary constants.

\subsection{Two-dimensional reductions leading to simpler PDE systems}

Below we briefly describe three two-dimensional reductions that lead to simpler PDE systems than \eqref{eq04ab}.

$1^\circ$. \textit{Solutions periodic in time $t$}.
System \eqref{eq04ab} has time-periodic solutions of the form
\begin{equation}
r=r(x,y),\quad \ \varphi=At+\xi(x,y),
\label{new07}
\end{equation}
where $A$ is an arbitrary constant, and the functions $r$ and $\xi$ are described by a two-dimensional PDE system, which is omitted here.
Solutions \eqref{new07} are another form of solutions \eqref{new05} obtained by going from Cartesian coordinates to polar ones.

$2^\circ$. \textit{Solutions periodic with respect to the angular variable}.
System \eqref{eq04ab} admits periodic solutions with respect to the angular variable of the form
\begin{equation}
r=r(\rho,t),\quad \ \varphi=A\theta+\eta(\rho,t),
\label{new08}
\end{equation}
where $A$ is an arbitrary constant, and the functions $r$ and $\eta$ are described by a two-dimensional PDE system, which is omitted here.

Note that the previously considered exact solutions \eqref{nn20} and \eqref{nn23}, which are described by ODEs, are special cases of solutions \eqref{new08}.

$3^\circ$. \textit{Solutions based on the use of a traveling wave type variable}.
System \eqref{eq04ab} also allows more complex solutions of the form
\begin{equation}
r=r(\rho,z),\quad \varphi=C_3t+C_4\theta+\zeta(\rho,z),\quad \ z=C_1t+C_2\theta,
\label{new09}
\end{equation}
where $C_1$, $C_2$, $C_3$, and $C_4$ are arbitrary constants,
and the functions $r$ and $\zeta$ are described by a two-dimensional PDE system, which is omitted here.

Solutions \eqref{new09} include as special cases the solutions \eqref{new07} and \eqref{new08}.

\section{Conclusion}
The nonlinear nonstationary Schr\"odinger equations of general form with two spatial variables, the dispersion and potential of which are given by one or two arbitrary functions, are investigated. Some solutions of these equations, which are expressed in quadratures or elementary functions, are found. Two-dimensional and one-dimensional reductions are described, which reduce the considered nonlinear PDEs to simpler equations of lower dimension or ordinary differential equations or systems of such equations. The obtained solutions can be used as test problems for numerical methods of integrating nonlinear PDEs of mathematical physics.

\section*{Acknowledgements}

The author thanks Prof. Nikolay A. Kudryashov for useful discussions.

The study was supported by the Ministry of Education and Science of the Russian Federation
within the framework of the state assignments No. 124012500440-9 and No. FSWU-2023-0031.


\renewcommand{\refname}{References}

\end{document}